# Spherical nano-inhomogeneity with the Steigmann–Ogden interface model under general uniform far-field stress loading

Junbo Wang[1], Peng Yan[1]*, Leiting Dong[1*], Satya N. Atluri [2]

[1]School of Aeronautic Science and Engineering, Beihang University, Beijing, CHINA

[2] Department of Mechanical Engineering, Texas Tech University, USA

**Abstract:** An explicit solution, considering the interface bending resistance as described by the Steigmann–Ogden interface model, is derived for the problem of a spherical nano-inhomogeneity (nanoscale void/inclusion) embedded in an infinite linear-elastic matrix under a general uniform far-field-stress (including tensile and shear stresses). The Papkovich-Neuber (P-N) general solutions, which are expressed in terms of spherical harmonics, are used to derive the analytical solution. A superposition technique is used to overcome the mathematical complexity brought on by the assumed interfacial residual stress in the Steigmann-Ogden interface model. Numerical examples show that the stress field, considering the interface bending resistance as with the Steigmann–Ogden interface model, differs significantly from that considering only the interface stretching resistance as with the Gurtin–Murdoch interface model. In addition to the size-dependency, another interesting phenomenon is observed: some stress components are invariant to interface bending stiffness parameters along a certain circle in the inclusion/matrix. Moreover, a characteristic line for the interface bending stiffness parameters is presented, near which the stress concentration becomes quite severe. Finally, the derived analytical solution with the Steigmann–Ogden interface model is provided in the supplemental MATLAB code, which can be easily executed, and used as a benchmark for semi-analytical solutions and numerical solutions in future studies.

* Corresponding author: ltdong@buaa.edu.cn (L. Dong). Address: School of Aeronautic Science and Engineering, Beihang University, Beijing, 100191, CHINA.

*Corresponding author: yanpeng117@buaa.edu.cn (P. Yan). Address: School of Aeronautic Science and Engineering, Beihang University, Beijing, 100191, CHINA.

## 1. Introduction

The "interface-stress" theory has attracted much attention due to its applicability to nanocomposites and nanostructured materials. The concept of interface stress was first introduced by Gibbs (1906) and has been extensively investigated since Gurtin and Murdoch (1975, 1978) incorporated it into continuum mechanics. In the Gurtin-Murdoch model, the interface is considered as a negligibly thin layer adhering to bulk materials without slipping, which only has stretching resistance but no bending resistance. Gurtin et al. (1998) generalized the original model by allowing all the components of the displacement vector to undergo a jump across the interface. The Gurtin–Murdoch model has been used to study nanosized rod (Altenbach et al., 2013; Grekov and Kostyrko, 2016), beams (Miller and Shenoy, 2000a; Eltaher et al., 2013; Ansari et al., 2015; Youcef et al., 2018), plates (Eremeyev et al., 2009; Ansari and Sahmani, 2011; Altenbach et al., 2012; Ansari and Norouzzadeh, 2016), shells (Altenbach et al., 2010; Altenbach and Eremeyev, 2011; Rouhi et al., 2016; Sahmani et al., 2016), films (Lu et al., 2011; Zhao and Rajapakse, 2013), wires (Diao et al., 2003; He and Lilley, 2008; Yvonnet et al., 2011), and inhomogeneities (Sharma et al., 2003; Duan et al., 2005a, b; Duan et al., 2005c; He and Li, 2006; Lim et al., 2006; Kushch et al., 2011; Kushch et al., 2013; Mi and Kouris, 2014; Nazarenko et al., 2016; Chen et al., 2018; Wang et al., 2018a), and much progress has been made in both analytical methods (Duan et al., 2009; Altenbach et al., 2013; Kushch et al., 2013; Dong et al., 2018) and numerical methods (Tian and Rajapakse, 2007; Feng et al., 2010; Dong and Pan, 2011).

In the Gurtin-Murdoch model, the surface/interface energy only depends on the surface/interface strains and the residual surface stress; thus the material interfaces are assumed to have only stretching resistance but no bending resistance. This makes the Gurtin–Murdoch model to be unable to account for the experimental observations and computational results on the size-dependence of the surface stresses for nanowires (McDowell et al., 2008; Yun and Park, 2009), nanoplates (Miller and Shenoy, 2000b) and nanoparticles (Medasani et al., 2007), since the elastic energy of the surface caused by curvature is neglected in this model. Steigmann and Ogden pointed out that the membrane in the Gurtin–Murdoch model cannot support compressive stress states (Ogden et al., 1997; Steigmann and Ogden, 1999); and thus cannot simulate surface features characterized by compressive surface stresses of any magnitude such as surface winkling and roughening. In order to overcome such deficiencies, Steigmann and Ogden (1999) generalized the Gurtin–Murdoch model to take into account both stretching and bending resistance of the membrane. The variational framework for the derivations of the basic equations for the model can be found in (Eremeyev and Lebedev, 2016; Zemlyanova and Mogilevskaya, 2018a).

In contrast to the large number of available studies for the Gurtin-Murdoch model, most of the literature on the Steigmann–Ogden model is focused on simple geometries, such as nanobeams (Chhapadia et al., 2011; Manav et al., 2018), nanowires (Zhao et al., 2015), rigid stamps (Zemlyanova, 2018a; Zemlyanova, 2018b), thin films (Ogden et al., 1997; Dryburgh and Ogden, 1999) and half-space materials (Li and Mi, 2018; Mi, 2018). The literature on nano-porous materials and nano-particle reinforced composites considering the Steigmann–Ogden surface elasticity model are rather limited (Gharahi and Schiavone, 2018; Han et al., 2018; Zemlyanova and Mogilevskaya, 2018a; Zemlyanova and Mogilevskaya, 2018b), and most of these studies are focused on 2D nano-inhomogeneity problems. Nevertheless, these studies on nano-inhomogeneities have shown that the interface bending resistance can significantly change the local stress distributions as well as the overall properties of nano-composites, and thus it should not be neglected. However, due to the mathematical complexity, studies for 3D nano-inhomogeneities based on the Steigmann-Ogden interface model under general uniform remote loading, have not been reported to the best of our knowledge. Especially, an analytical solution for a spherical nano-inhomogeneity, which can serve as the benchmark for numerical and semi-analytical solutions, is desirable.

In this study, an analytical solution considering the interface bending resistance based on the Steigmann-Ogden (S-O) interface model is derived for the first time, for a spherical nano-inhomogeneity (nanoscale void/inclusion) embedded in an infinite matrix under general uniform far-field loading. The Papkovich-Neuber (P-N) general solutions are used, together with spherical harmonics to derive the solution of a spherical nano-inhomogeneity with S-O interface, embedded in an infinite matrix, under general uniform far-field loading. This approach was previously used to develop a series of novel numerical tools named as "computational grains" (Dong and Atluri, 2012a, b; Wang et al., 2018c), for direct numerical simulation of microstructures with a large number of heterogeneities, considering different shapes, distributions, constitutive relations, physics, and interfaces. For example, it was used in (Dong and Atluri, 2012a, b), in which computational grains (mathematical or virtual finite-sized domains of polyhedral geometries, each with embedded spherical or ellipsoidal inclusions/voids) were developed for highly efficient direct numerical simulation of the micromechanics of composites. It was also used to deal with composites with coated spherical inclusions or fiber reinforcements (Wang et al., 2018b; Wang et al., 2018c). One may also follow the procedures presented in this paper to solve other problems of inhomogeneities with different interface models and different shapes, e.g. cylindrical nano-inhomogeneities and ellipsoidal nano-inhomogeneities, by expressing the Papkovich-Neuber potentials in different types of harmonics (cylindrical, ellipsoidal, etc.). More complex loads may also be considered, such as far-field bending.

The rest of this paper is organized as follows: In Section 2, the governing equations for the 3D nano-inhomogeneity with Steigmann-Ogden interface are briefly stated. In Section 3, the Papkovich-Neuber solutions and spherical harmonics are detailed. Then using the Steigmann-Ogden interface description and the far-field conditions, the explicit analytical solution to the considered nano-inhomogeneity problem is given in Section 4. In Section 5, we discuss the influences of the interface bending on stress distributions within and around the nano-inhomogeneity(nano-void/inclusion), when the far-field tensile/shear loads are applied. In Section 6, we complete this paper with some concluding remarks.

## 2. The governing linear elasticity equations

The problem of a nano-inhomogeneity embedded in an infinite elastic matrix subjected to general uniform far-field stress loading is considered, as shown in Fig. 1. Solutions of 3D linear elasticity for the matrix and the inhomogeneity should satisfy the equations of stress equilibrium, strain displacement-gradient compatibility, as well as the constitutive relations in each domain $\Omega^j$:

$$\nabla \cdot \boldsymbol{\sigma}^j + \mathbf{f}^j = 0 \tag{1}$$

$$\boldsymbol{\varepsilon}^j = \frac{1}{2}(\nabla \mathbf{u}^j + (\nabla \mathbf{u}^j)^T) \tag{2}$$

$$\boldsymbol{\sigma}^j = \lambda^j \mathrm{tr}(\boldsymbol{\varepsilon}^j)\mathbf{I}_3 + 2\mu^j \boldsymbol{\varepsilon}^j \tag{3}$$

where the superscript $j = \mathrm{m}$ denotes the matrix, and $j = \mathrm{i}$ denotes the inhomogeneity. $\boldsymbol{\sigma}^j, \boldsymbol{\varepsilon}^j, \mathbf{u}^j$ are stresses, strains, and displacements in matrix/inhomogeneity. $\mathbf{f}^j$ is the body force which can be neglected here. ($\nabla \cdot$) and $\nabla$ are the divergence and gradient operators, respectively. $\lambda^j = \dfrac{\nu^j E^j}{(1-2\nu^j)(1+\nu^j)}$ and $\mu^j = \dfrac{E^j}{2(1+\nu^j)}$ are Lamé constants, where $E^j$ and $\nu^j$ are the Youngs modulus and Poissons ratio, respectively. $\mathbf{I}_3$ is the 3D unit tensor and $\mathbf{I}_3 = \mathbf{e}_r \otimes \mathbf{e}_r + \mathbf{e}_\theta \otimes \mathbf{e}_\theta + \mathbf{e}_\varphi \otimes \mathbf{e}_\varphi$ in spherical coordinates, where $\mathbf{e}_r$, $\mathbf{e}_\theta$, $\mathbf{e}_\varphi$ are base vectors. $\mathrm{tr}(\boldsymbol{\varepsilon}^j)$ denotes the trace of the strain tensor.

A general uniform far-field stress loading, with arbitrary combinations of shears and tensions, and can be written as:

$$\boldsymbol{\sigma}^m = \boldsymbol{\sigma}^0 \quad \text{at infinity} \tag{4}$$

$\boldsymbol{\sigma}^0 = \begin{pmatrix} \sigma^0_{xx} & \sigma^0_{xy} & \sigma^0_{xz} \\ \sigma^0_{xy} & \sigma^0_{yy} & \sigma^0_{yz} \\ \sigma^0_{xz} & \sigma^0_{yz} & \sigma^0_{zz} \end{pmatrix}$, which includes 6 independent components of normal stresses and shear stresses in general.

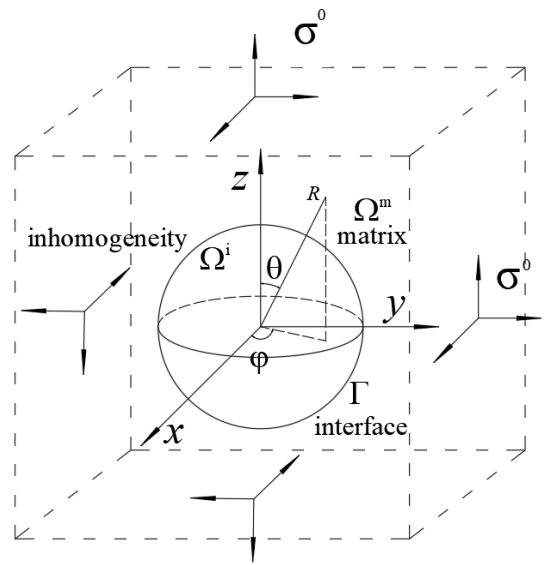

Fig. 1. A spherical inhomogeneity embedded in an infinite matrix under far-field loading

The governing equations for the S–O interface model can be obtained by taking the first variation of the following functional $\Pi$ :

$$\Pi(\mathbf{u}^m,\mathbf{u}^i) = \sum_{j=m,i}\int_{\Omega^j} W^j d\Omega + \int_\Gamma U^s dS - \int_{S_t} \bar{\mathbf{t}} \cdot \mathbf{u}^m dS - \sum_{j=m,i}\int_{\Omega^j} \mathbf{f}^j \cdot \mathbf{u}^j d\Omega \qquad (5)$$

where

$$W^j = \frac{1}{2}\boldsymbol{\sigma}^j : \boldsymbol{\varepsilon}^j \qquad (6)$$

$$U^s = (\mu^s - \sigma^s)\boldsymbol{\varepsilon}^s : \boldsymbol{\varepsilon}^s + \frac{1}{2}(\lambda^s + \sigma^s)\mathrm{tr}(\boldsymbol{\varepsilon}^s)^2 + \chi^s \boldsymbol{\kappa}^s : \boldsymbol{\kappa}^s + \frac{1}{2}\zeta^s \mathrm{tr}(\boldsymbol{\kappa}^s)^2$$
$$+ \sigma^s (1 + \mathrm{tr}(\boldsymbol{\varepsilon}^s)) + \frac{1}{2}\sigma^s (\nabla_s \mathbf{u}^s):(\nabla_s \mathbf{u}^s) \qquad (7)$$

$$\boldsymbol{\varepsilon}^s = \frac{1}{2}\left[\nabla_s \mathbf{u}^s \cdot \mathbf{I}_s + \mathbf{I}_s \cdot (\nabla_s \mathbf{u}^s)^T\right] \qquad (8)$$

$$\boldsymbol{\kappa}^s = -\frac{1}{2}\left[\nabla_s \vartheta \cdot \mathbf{I}_s + \mathbf{I}_s \cdot (\nabla_s \vartheta)^T\right] \qquad (9)$$

$$\vartheta = \nabla_s (\mathbf{n} \cdot \mathbf{u}^s) + \mathbf{B} \cdot \mathbf{u}^s \qquad (10)$$

$$\mathbf{B} = -\nabla_s \mathbf{n} \qquad (11)$$

where $\mathbf{u}^s, \boldsymbol{\varepsilon}^s$ and $\boldsymbol{\kappa}^s$ are interface displacement, strain and curvature, respectively. $\lambda^s$ and $\mu^s$ are the interface Lamé constants characterizing the interface stretching. $\chi^s$ and $\zeta^s$ are stiffness parameters characterizing the interface bending. $\sigma^s$ is the residual surface stress. $\mathbf{I}_s$ is the unit tangent tensor defined on the interface and $\mathbf{I}_s = \mathbf{e}_\theta \otimes \mathbf{e}_\theta + \mathbf{e}_\varphi \otimes \mathbf{e}_\varphi$ in spherical coordinates. $\nabla_s = (\mathbf{I}_3 - \mathbf{nn}) \cdot \nabla$ is the gradient operator defined on the interface where $\mathbf{n}$ is the unit outer-normal vector of the interface $\Gamma$. $\bar{\mathbf{t}}$ is the prescribed boundary tractions at the traction boundary $S_t$.

It should be pointed out that functional $\Pi$ is the total potential energy of nano-composites, which is the sum of the elastic strain energy of bulk materials, the interface energy and the potential energy associated to the applied forces. In Eq.(5), $\mathbf{u}^j (j = m \text{ or } i)$ satisfies the compatibility and constitutive equations a-priori, and

$$\mathbf{u}^m = \mathbf{u}^i \text{ at } \Gamma \qquad (12)$$

$$\mathbf{u}^m = \bar{\mathbf{u}} \text{ at } S_u \tag{13}$$

where $\bar{\mathbf{u}}$ is the prescribed boundary displacements at the displacement boundary $S_u$. The Euler-Lagrange equations of Eq.(5) are the equilibrium equation(Eq.(1)), the traction boundary conditions ($\boldsymbol{\sigma}^m \cdot \mathbf{n} = \bar{\mathbf{t}}$ at $S_t$) and the stress jump across the interface:

$$\mathbf{n} \cdot \Delta \boldsymbol{\sigma} = \nabla_s \cdot \left[ \boldsymbol{\tau}^s + \left( \nabla_s \cdot \mathbf{m}^s \right) \mathbf{n} \right] - \left( \nabla_s \cdot \mathbf{n} \right) \mathbf{n} \cdot \left( \nabla_s \cdot \mathbf{m}^s \right) \mathbf{n} \quad \text{at } \Gamma \tag{14}$$

where $\boldsymbol{\tau}^s$ and $\mathbf{m}^s$ are interface stress and bending moment, respectively:

$$\boldsymbol{\tau}^s = \sigma^s \mathbf{I}_s + 2(\mu^s - \sigma^s)\boldsymbol{\varepsilon}^s + (\lambda^s + \sigma^s)\text{tr}(\boldsymbol{\varepsilon}^s)\mathbf{I}_s + \sigma^s \nabla_s \mathbf{u}^s \tag{15}$$

$$\mathbf{m}^s = 2\chi^s \boldsymbol{\kappa}^s + \zeta^s \text{tr}(\boldsymbol{\kappa}^s)\mathbf{I}_s \tag{16}$$

Eq.(14) was first presented in Eremeyev and Lebedev (2016), and was generalized by Zemlyanova and Mogilevskaya (2018a) to further take the interface residual tension into consideration. With these two papers, one can find the detailed derivation.

## 3. Papkovich-Neuber solutions with spherical harmonics

### 3.1. Papkovich-Neuber solutions

In order to solve the governing equations Eqs. (1-3), Navier's equation

$$(\lambda^j + \mu^j)\nabla(\nabla \cdot \mathbf{u}^j) + \mu^j \nabla^2 \mathbf{u}^j + \mathbf{f}^j = 0 \tag{17}$$

is derived (Lurie, 2005). The solutions of Navier's equation can be represented in the form of harmonic functions (Papkovich, 1932; Neuber, 1934; Lurie, 2005) when the body force is neglected:

$$\mathbf{u}^j = [4(1-v^j)\mathbf{B}^j - \nabla(\mathbf{R} \cdot \mathbf{B}^j + B_0^j)]/2\mu^j \tag{18}$$

where $B_0^j$ and $\mathbf{B}^j = \begin{bmatrix} B_1^j & B_2^j & B_3^j \end{bmatrix}^T$ are scalar and vector harmonic functions. $\mathbf{R}$ is the position vector.

According to Slobodyansky (1954), by dropping $B_0^k$ the following solution:

$$\mathbf{u}^j = [4(1-v^j)\mathbf{B}^j - \nabla \mathbf{R} \cdot \mathbf{B}^j]/2\mu^j \tag{19}$$

is complete for an infinite domain external to a closed surface, thus will be applied in the matrix. However, for a simply-connected domain, Eq.(19) is incomplete when $v^j = 0.25$. Therefore, another general solution,

$$\mathbf{u}^j = \left[ 4(1-v^j)\mathbf{B}^j + \mathbf{R} \cdot \nabla \mathbf{B}^j - \mathbf{R}\nabla \cdot \mathbf{B}^j) \right]/2\mu^j \tag{20}$$

is obtained by expressing $B_0^j$ in Eq.(18) as a specific function of $\mathbf{B}^j$. This general solution is complete for any Poisson ratio $v^j$ in a simply connected domain, thus will be applied in the inclusion.

### 3.2. Spherical harmonics

The displacement field in the inclusion can be derived by substituting the non-singular harmonics:

$$\mathbf{B}_p^i = \sum_{n=0}^{\infty} R^n \left\{ \mathbf{a}_0^0 YC_0^0(\theta,\varphi) + \sum_{l=1}^{n} \left[ \mathbf{a}_n^l YC_n^l(\theta,\varphi) + \mathbf{b}_n^l YS_n^l(\theta,\varphi) \right] \right\} \tag{21}$$

into Eq.(20):

$$\mathbf{u}^i = \mathbf{u}_p^i = \left[ 4(1-v^i)\mathbf{B}_p^i + \mathbf{R} \cdot \nabla \mathbf{B}_p^i - \mathbf{R}\nabla \cdot \mathbf{B}_p^i) \right]/2\mu^i \tag{22}$$

where $\mathbf{a}_n^l$, $\mathbf{b}_n^l$ are the unknown coefficients to be determined. $YC_n^l(\theta,\varphi)$ and $YS_n^l(\theta,\varphi)$ are spherical harmonics:

$$YC_n^l(\theta,\varphi) = \sqrt{\frac{2n+1}{4\pi}\frac{(n-l)!}{(n+l)!}}P_n^l(\cos(\theta))\cos(k\varphi)$$

$$YS_n^l(\theta,\varphi) = \sqrt{\frac{2n+1}{4\pi}\frac{(n-l)!}{(n+l)!}}P_n^l(\cos(\theta))\sin(k\varphi) \quad (23)$$

$$P_n^l(x) = \frac{(1-x^2)^{\frac{l}{2}}}{2^n n!}\frac{d^{n+l}}{dx^{n+l}}(x^2-1)^n$$

The displacement field in the matrix is the summation of $\mathbf{u}_p^m$ (the non-singular part) and $\mathbf{u}_k^m$ (the singular part, with the singularity located at the centre of the inclusion). $\mathbf{u}_p^m$ can be derived by substituting

$$\mathbf{B}_p^m = \sum_{n=0}^{\infty} R^n \left\{ \mathbf{c}_0^0 YC_0^0(\theta,\varphi) + \sum_{l=1}^{n}\left[\mathbf{c}_n^l YC_n^l(\theta,\varphi) + \mathbf{d}_n^l YS_n^l(\theta,\varphi)\right] \right\} \quad (24)$$

into Eq.(20), and $\mathbf{u}_k^m$ can be derived by substituting

$$\mathbf{B}_k^m = \sum_{n=0}^{\infty} R^{-(n+1)} \left\{ \mathbf{s}_0^0 YC_0^0(\theta,\varphi) + \sum_{l=1}^{n}\left[\mathbf{s}_n^l YC_n^l(\theta,\varphi) + \mathbf{t}_n^l YS_n^l(\theta,\varphi)\right] \right\} \quad (25)$$

into Eq. (19):

$$\begin{aligned}
\mathbf{u}^m &= \mathbf{u}_p^m + \mathbf{u}_k^m \\
\mathbf{u}_p^m &= \left[4(1-\nu^m)\mathbf{B}_p^m + \mathbf{R}\cdot\nabla\mathbf{B}_p^m - \mathbf{R}\nabla\cdot\mathbf{B}_p^m)\right]/2\mu^m \\
\mathbf{u}_k^m &= [4(1-\nu^m)\mathbf{B}_k^m - \nabla\mathbf{R}\cdot\mathbf{B}_k^m]/2\mu^m
\end{aligned} \quad (26)$$

where $\mathbf{s}_n^l, \mathbf{t}_n^l, \mathbf{c}_n^l, \mathbf{d}_n^l$ are the unknown coefficients.

## 4. Solution to the problem by a superposition technique

By employing Papkovitch–Neuber solutions, the elastic field resulting from the general uniform far-field stress loading is obtained explicitly. First, we consider the case that the remote loading has only one non-zero stress component $\sigma_{xx}^0$. Because $\mathbf{u}^j$ ($j = $ m or c) satisfies Navier's equation a priori, the unknown coefficients are determined by enforcing the far-field boundary condition (Eq.(4)) and interface conditions (Eq.(12) and Eq.(14)). After solving the unknown coefficients, the displacement field $\mathbf{u}_{xx}^{j}$ (here we use $\mathbf{u}_{st}^{j}$ to denote the displacement field when the remote loading has only one non-zero component $\sigma_{st}^{0}$) can be obtained:

$$u_{rxx}^{m} = M_{1xx}r + \frac{M_{2xx}}{r^2} - \frac{1}{8}\left(2M_{3xx}r + \frac{2(5-4v^m)}{r^2}M_{4xx} - 3\frac{M_{5xx}}{r^4}\right)(1+$$

$$3\cos[2\theta] - 6\cos[2\phi]\sin[\theta]^2)$$

$$u_{\theta xx}^{m} = 3\left(M_{3xx}r + \frac{2(1-2v^m)}{r^2}M_{4xx} + \frac{M_{5xx}}{r^4}\right)\cos[\theta]\cos[\phi]^2\sin[\theta];$$

$$u_{\phi xx}^{m} = -3\left(M_{3xx}r + \frac{2(1-2v^m)}{r^2}M_{4xx} + \frac{M_{5xx}}{r^4}\right)\cos[\phi]\sin[\theta]\sin[\phi]; \quad (27)$$

$$u_{rxx}^{i} = C_{1xx}r - \frac{1}{8}\left(12v^i C_{2xx}r^3 + 2C_{3xx}r\right)\left(1+3\cos[2\theta] - 6\cos[2\phi]\sin[\theta]^2\right);$$

$$u_{\theta xx}^{i} = 3\left((7-4v^i)C_{2xx}r^3 + C_{3xx}r\right)\cos[\theta]\cos[\phi]^2\sin[\theta];$$

$$u_{\phi xx}^{i} = -3\left((7-4v^i)C_{2xx}r^3 + C_{3xx}r\right)\cos[\phi]\sin[\theta]\sin[\phi]$$

By using the same procedure, the displacement field $\mathbf{u}_{yy}^{k}$ with the remote tensile stress $\sigma_{yy}^{0}$ can be written as:

$$u_{ryy}^{m} = M_{1yy}r + \frac{M_{2yy}}{r^2} - \frac{1}{8}\left(2M_{3yy}r + \frac{2(5-4v^m)}{r^2}M_{4yy} - 3\frac{M_{5yy}}{r^4}\right)(1+$$

$$3\cos[2\theta] + 6\cos[2\phi]\sin[\theta]^2);$$

$$u_{\theta yy}^{m} = 3\left(M_{3yy}r + \frac{2(1-2v^m)}{r^2}M_{4yy} + \frac{M_{5yy}}{r^4}\right)\cos[\theta]\sin[\theta]\sin[\phi]^2;$$

$$u_{\phi yy}^{m} = 3\left(M_{3yy}r + \frac{2(1-2v^m)}{r^2}M_{4yy} + \frac{M_{5yy}}{r^4}\right)\cos[\phi]\sin[\theta]\sin[\phi]; \quad (28)$$

$$u_{ryy}^{i} = C_{1yy}r - \frac{1}{8}\left(12v^i C_{2yy}r^3 + 2C_{3yy}r\right)\left(1+3\cos[2\theta] + 6\cos[2\phi]\sin[\theta]^2\right);$$

$$u_{\theta yy}^{i} = 3\left((7-4v^i)C_{2yy}r^3 + C_{3yy}r\right)\cos[\theta]\sin[\theta]\sin[\phi]^2;$$

$$u_{\phi yy}^{i} = 3\left((7-4v^i)C_{2yy}r^3 + C_{3yy}r\right)\cos[\phi]\sin[\theta]\sin[\phi]$$

The displacement field $\mathbf{u}_{zz}^{k}$ with the remote tensile stress $\sigma_{zz}^{0}$ can be written as:

$$u_{rzz}^m = M_{1zz}r + \frac{M_{2zz}}{r^2} + \frac{1}{4}\left(2M_{3zz}r + \frac{2(5-4v^m)}{r^2}M_{4zz} - 3\frac{M_{5zz}}{r^4}\right)(1+3\cos[2\theta]);$$

$$u_{\theta zz}^m = -3\left(M_{3zz}r + \frac{2(1-2v^m)}{r^2}M_{4zz} + \frac{M_{5zz}}{r^4}\right)\cos[\theta]\sin[\theta];$$

$$u_{\phi zz}^m = 0;$$

$$u_{rzz}^i = C_{1zz}r + \frac{1}{4}\left(12v^i C_{2zz}r^3 + 2C_{3zz}r\right)(1+3\cos[2\theta]);$$

$$u_{\theta zz}^i = -3\left((7-4v^i)C_{2zz}r^3 + C_{3zz}r\right)\cos[\theta]\sin[\theta];$$

$$u_{\phi zz}^i = 0$$

(29)

The displacement field $\mathbf{u}_{xy}^k$ with the remote shear stress $\sigma_{xy}^0$ can be written as:

$$u_{rxy}^m = \frac{M_{2xy}}{r^2} + \frac{3}{2}\left(2M_{3xy}r + \frac{2(5-4v^m)}{r^2}M_{4xy} - 3\frac{M_{5xy}}{r^4}\right)\sin[\theta]\sin[\theta]\sin[2\phi];$$

$$u_{\theta xy}^m = \frac{3}{2}\left(M_{3xy}r + \frac{2(1-2v^m)}{r^2}M_{4xy} + \frac{M_{5xy}}{r^4}\right)\sin[2\theta]\sin[2\phi];$$

$$u_{\phi xy}^m = 3\left(M_{3xy}r + \frac{2(1-2v^m)}{r^2}M_{4xy} + \frac{M_{5xy}}{r^4}\right)\sin[\theta]\cos[2\phi];$$

$$u_{rxy}^i = C_{1xy}r + \frac{3}{2}\left(12v^i C_{2xy}r^3 + 2C_{3xy}r\right)\sin[\theta]\sin[\theta]\sin[2\phi];$$

$$u_{\theta xy}^i = \frac{3}{2}\left((7-4v^i)C_{2xy}r^3 + C_{3xy}r\right)\sin[2\theta]\sin[2\phi];$$

$$u_{\phi xy}^i = 3\left((7-4v^i)C_{2xy}r^3 + C_{3xy}r\right)\sin[\theta]\cos[2\phi]$$

(30)

The displacement field $\mathbf{u}_{yz}^k$ with the remote shear stress $\sigma_{yz}^0$ can be written as:

$$u_{ryz}^m = \frac{M_{2yz}}{r^2} + \frac{3}{2}\left(2M_{3yz}r + \frac{2(5-4v^m)}{r^2}M_{4yz} - 3\frac{M_{5yz}}{r^4}\right)\sin[2\theta]\sin[\phi];$$

$$u_{\theta yz}^m = 3\left(M_{3yz}r + \frac{2(1-2v^m)}{r^2}M_{4yz} + \frac{M_{5yz}}{r^4}\right)\cos[2\theta]\sin[\phi];$$

$$u_{\phi yz}^m = 3\left(M_{3yz}r + \frac{2(1-2v^m)}{r^2}M_{4yz} + \frac{M_{5yz}}{r^4}\right)\cos[\theta]\cos[\phi];$$

$$u_{ryz}^i = C_{1yz}r + \frac{3}{2}\left(12v^i C_{2yz}r^3 + 2C_{3yz}r\right)\sin[2\theta]\sin[\phi];$$

$$u_{\theta yz}^i = 3\left((7-4v^i)C_{2yz}r^3 + C_{3yz}r\right)\cos[2\theta]\sin[\phi];$$

$$u_{\phi yz}^i = 3\left((7-4v^i)C_{2yz}r^3 + C_{3yz}r\right)\cos[\theta]\cos[\phi]$$

(31)

The displacement field $\mathbf{u}_{zx}^k$ with the remote shear stress $\sigma_{zx}^0$ can be written as:

$$u_{rzx}^{m} = \frac{M_{2zx}}{r^2} + \frac{3}{2}\left(2M_{3zx}r + \frac{2(5-4v^m)}{r^2}M_{4zx} - 3\frac{M_{5zx}}{r^4}\right)\sin[2\theta]\cos[\phi];$$

$$u_{\theta zx}^{m} = 3\left(M_{3zx}r + \frac{2(1-2v^m)}{r^2}M_{4zx} + \frac{M_{5zx}}{r^4}\right)\cos[2\theta]\cos[\phi];$$

$$u_{\phi zx}^{m} = -3\left(M_{3zx}r + \frac{2(1-2v^m)}{r^2}M_{4zx} + \frac{M_{5zx}}{r^4}\right)\cos[\theta]\sin[\phi]; \qquad (32)$$

$$u_{rzx}^{i} = C_{1zx}r + \frac{3}{2}\left(12v^i C_{2zx}r^3 + 2C_{3zx}r\right)\sin[2\theta]\cos[\phi];$$

$$u_{\theta zx}^{i} = 3\left((7-4v^i)C_{2zx}r^3 + C_{3zx}r\right)\cos[2\theta]\cos[\phi];$$

$$u_{\phi zx}^{i} = -3\left((7-4v^i)C_{2zx}r^3 + C_{3zx}r\right)\cos[\theta]\sin[\phi]$$

where $M_{pst}$ ($p=1,...,5$ and $s,t=x,y,z$) and $C_{qst}$ ($q=1,2,3$ and $s,t=x,y,z$) are constants given in Appendix A.

For the case that the remote loading is zero, the displacement field $\mathbf{u}_0^k$ can be written as:

$$u_{r0}^{m} = R^3(1-2v^i)\sigma^s / (r^2((-2+4v^i)\lambda^s - R(\mu^i + v^i\mu^i + 2\mu^m - 4v^i\mu^m)$$
$$+(-1+2v^i)(2\mu^s + \sigma^s)));$$
$$u_{\theta 0}^{m} = 0;$$
$$u_{\phi 0}^{m} = 0;$$
$$u_{r0}^{i} = r(-1+2v^i)\sigma^s / ((2-4v^i)\lambda^s + R(\mu^i + v^i\mu^i + 2\mu^m - 4v^i\mu^m) + 2 \qquad (33)$$
$$\mu^s + \sigma^s - 2v^i(2\mu^s + \sigma^s));$$
$$u_{\theta 0}^{i} = 0;$$
$$u_{\phi 0}^{i} = 0;$$

Now we have obtained the basic solutions for a spherical inhomogeneity under different remote loading cases. However, the analytical solution under general remote loading $\boldsymbol{\sigma}^0$ is not simply an additive combination of the above Eqs.(27-32), due to the existence of $\sigma^s \mathbf{I}_s$ in Eq.(15). If we simply add Eqs.(27-32) together, the interface stress will be:

$$\boldsymbol{\tau}^s = 6\sigma^s \mathbf{I}_s + 2(\mu^s - \sigma^s)\boldsymbol{\varepsilon}^s + (\lambda^s + \sigma^s)\text{tr}(\boldsymbol{\varepsilon}^s)\mathbf{I}_s + \sigma^s \nabla_s \mathbf{u}^s \qquad (34)$$

Obviously, the extra 5 terms of $\sigma^s \mathbf{I}_t$ should be eliminated, thus the analytical solution under general loading $\boldsymbol{\sigma}^0$ should be written as:

$$\mathbf{u}^j = \mathbf{u}^j_{xx} + \mathbf{u}^j_{yy} + \mathbf{u}^j_{zz} + \mathbf{u}^j_{xy} + \mathbf{u}^j_{yz} + \mathbf{u}^j_{zx} - 5\mathbf{u}^j_0 \quad j = m, i \qquad (35)$$

It can be easily proved that the displacement solution (Eq.(35)) given by the superposition satisfies the governing equations by substituting Eq.(35) into Eqs.(1-16). The solution Eq.(35) reveals that the interface effect is size dependent and different interface properties will influence the stress concentration. If the interface stress effect is neglected ($\lambda^s = 0$, $\mu^s = 0$, $\sigma^s = 0$, and $\zeta^s = 0$), Eq.(35) will reduce to the classical Eshelby solution. If the surface bending resistance is neglected ($\chi^s = 0$ and $\zeta^s = 0$), Eq.(35) can be degenerated into the solution considering the Gurtin-Murdoch interface model $\chi^s = 0$ (Duan et al., 2009; Mi and Kouris, 2014). This is also demonstrated in the numerical example shown in Fig.2.

Using strain displacement-gradient compatibility and the constitutive relations, the stress field can be obtained easily.

## 5. Results and discussion

In this section, we present some numerical examples to illustrate the contribution of interface elasticity with bending resistance. Due to the lack of experimental data, here hypothetical parameters are used to demonstrate the difference between the classical results and those for the Steigmann–Ogden model.

*5.1. A nano-void embedded in an infinite matrix*

The first case investigated is an infinite matrix containing a spherical void. The material properties for the matrix are: $E^m = 71$ GPa and $v^m = 0.35$. The interface elastic constants are: $\lambda^s = 3.4939$ N/m, $\mu^s = -5.4251$ N/m and $\sigma^s = 0.5689$ N/m (Tian, 2006). The radius of the nano-void is $R_{void} = 1$ nm.

In order to verify the analytical solution presented in this study, we compare the results given by Eq.(35) with those considering the G-M model (Duan et al., 2009; Mi and Kouris, 2014). In this example, we set $\chi^s = \zeta^s = 0$. Fig. 2. gives the comparison between the solutions given in (Duan et al., 2009; Mi and Kouris, 2014) and that in our study. The results show that Eq.(35) can be degenerated into the solutions considering the Gurtin-Murdoch interface model (Duan et al., 2009; Mi and Kouris, 2014) when the surface bending resistance is neglected. Eq.(35) can be further degenerated into the classical Eshelby solutions when the interface stress vanishes.

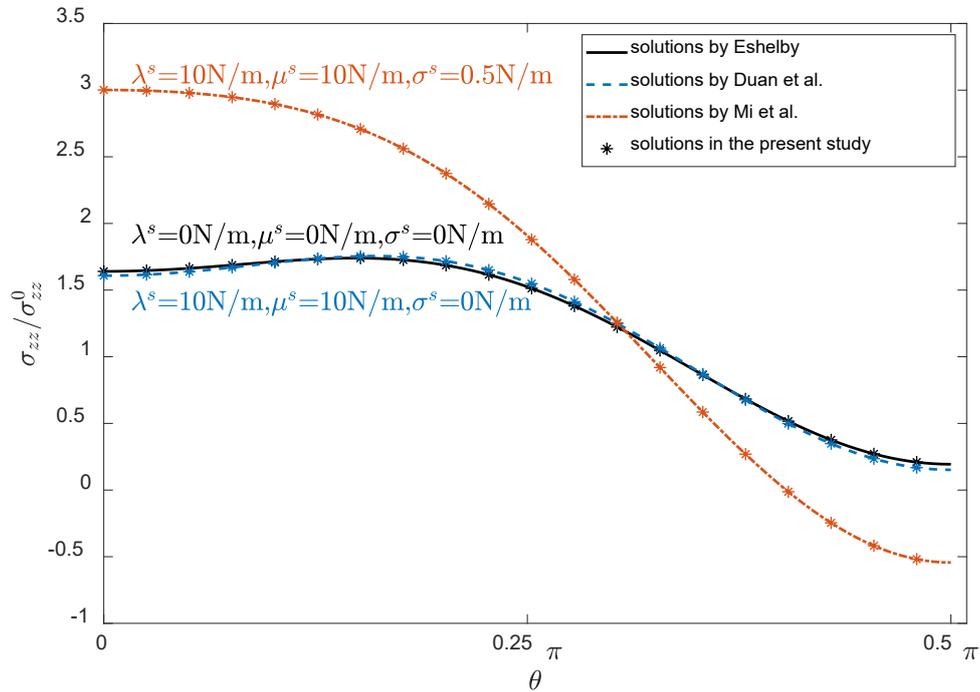

Fig. 2. The comparison between the analytical solution in (Duan et al., 2009; Mi and Kouris, 2014) and that in this study.

A series of parametric studies on interface properties are conducted to investigate the influence of surface bending resistance on stress distributions. Fig. 3. shows variation of stress components along the meridian line of the void with different interface bending stiffness parameters and different remote loading. From Fig. 3., we can see that the stress distributions are strongly affected by the interface bending stiffness parameters. Moreover, an interesting phenomenon is observed from Fig.3(a): $\sigma_{zz}$ with different bending stiffness parameters have the same value along the circle $\theta = \theta_{void}$ (the value of $\theta_{void}$ can be found in Appendix A) on the surface of the void, when $\lambda^s$, $\mu^s$ and $\sigma^s$ keep unchanged.

In addition, the influence of bending stiffness parameters on $\sigma_{zz}$ at the point $(0, 0, R_{void})$ is further studied, as shown in Fig.4. It is observed that as $(\chi^s, \zeta^s)$ gets close to the characteristic line:

$$5\chi^s + 3\zeta^s + c_{void} = 0 \tag{36}$$

the stress concentration becomes quite severe. The constant $c_{void}$ is also given in Appendix A.

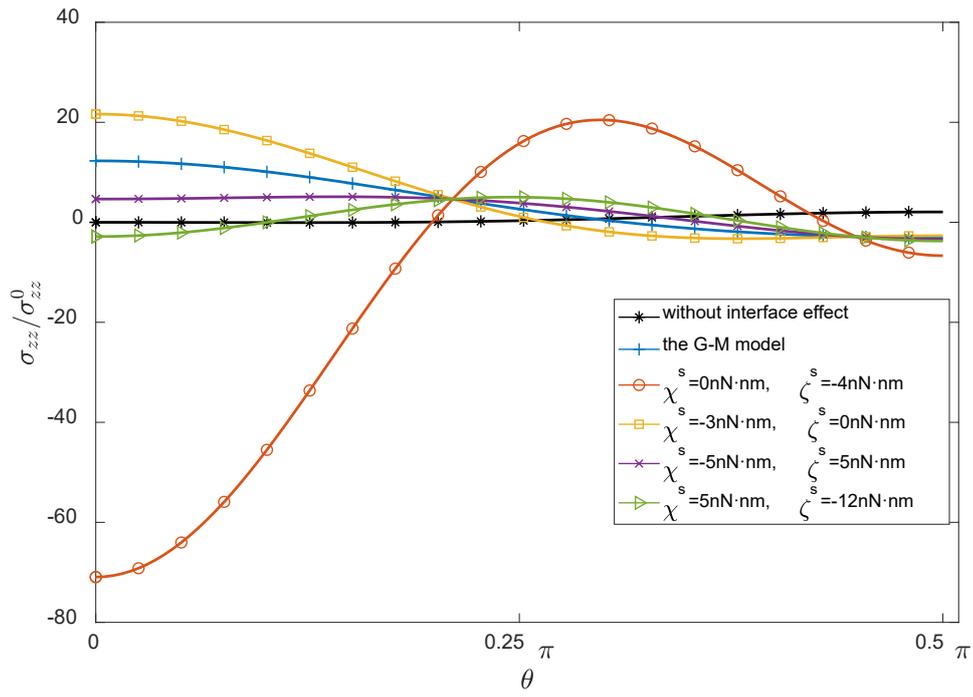

(a)

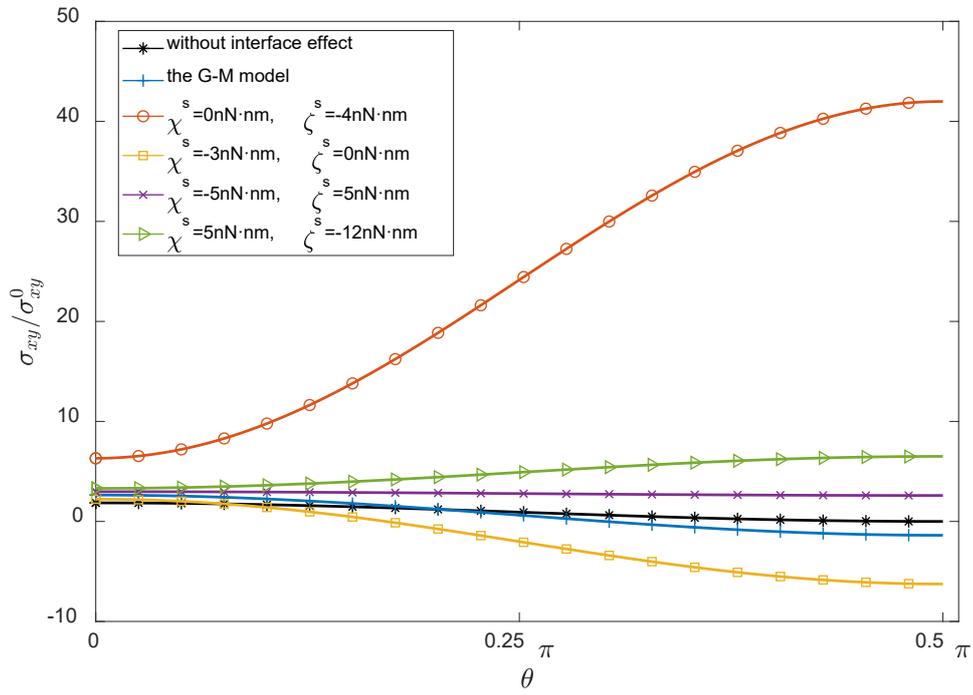

(b)

Fig. 3. (a) Variation of $\sigma_{zz}$ along the meridian line of the void for different interface

properties when $\boldsymbol{\sigma}^0 = \begin{pmatrix} 0 & 0 & 0 \\ 0 & 0 & 0 \\ 0 & 0 & 100 \end{pmatrix}$ MPa and (b) variation of $\sigma_{xy}$ along the meridian line of the void for different interface properties when $\boldsymbol{\sigma}^0 = \begin{pmatrix} 0 & 100 & 0 \\ 100 & 0 & 0 \\ 0 & 0 & 0 \end{pmatrix}$ MPa.

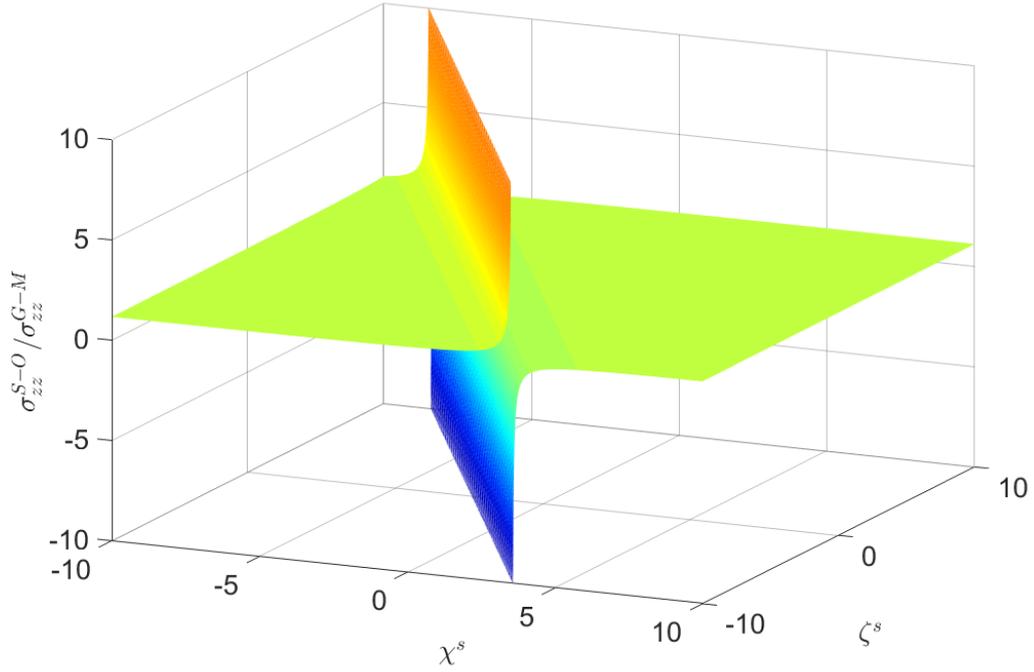

Fig. 4. The influence of $\chi^s$ and $\zeta^s$ on $\sigma_{zz}$ at the point $(0,0,R_{\text{void}})$ when $\boldsymbol{\sigma}^0 = \begin{pmatrix} 0 & 0 & 0 \\ 0 & 0 & 0 \\ 0 & 0 & 100 \end{pmatrix}$ MPa is applied. The stress from the Steigmann–Ogden interface model $\sigma_{zz}^{\text{S-O}}$ is normalized by that from the Gurtin-Murdoch interface model $\sigma_{zz}^{\text{G-M}}$.

Fig. 5. shows $\sigma_{zz}/\sigma_{zz}^0$ at point $(0,0,R_{\text{void}})$ with different interface bending stiffness parameters and with different void radiuses. The results reveal that the stress concentration is size dependent, and such a size-dependency is influenced by the interface bending stiffness parameters. The smaller the void is, the more significant interface effects are.

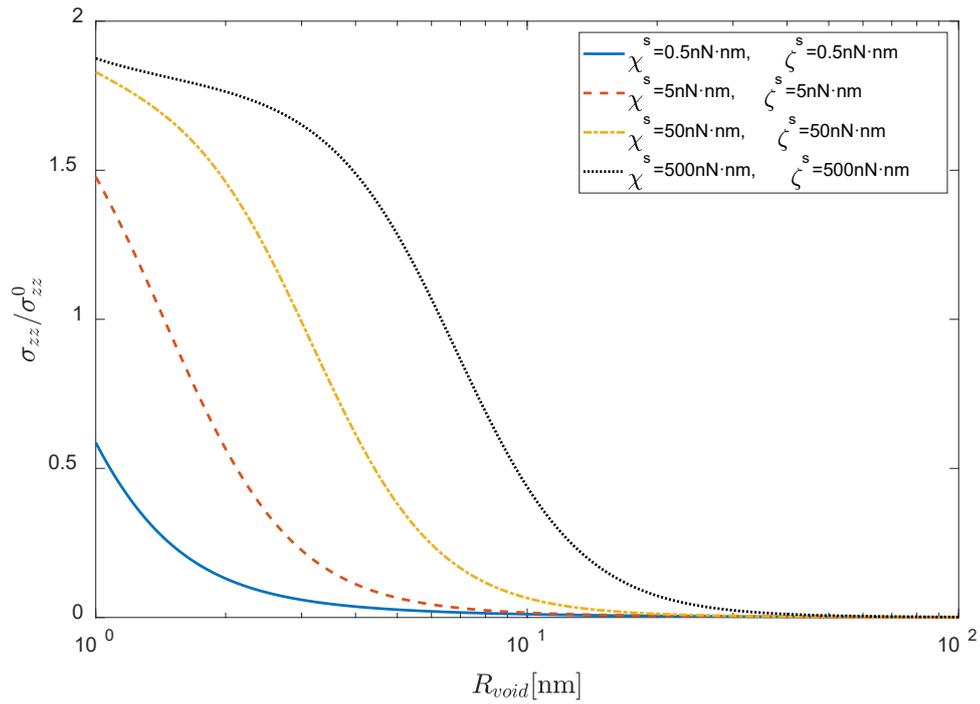

Fig. 5. Variation of $\sigma_{zz}/\sigma_{zz}^0$ at point $(0,0,R_{void})$ with the void radiuses for different interface properties.

### 5.2. A nano-inclusion embedded in an infinite matrix

The second case investigated is an infinite matrix containing a spherical nano-inclusion under general uniform far-field loading. The material properties for the inclusion are: $E^i = 410$ GPa and $v^i = 0.14$, while the material properties for the matrix are: $E^m = 71$ GPa and $v^m = 0.35$. The interface elastic constants are: $\lambda^s = 3.4939$ N/m, $\mu^s = -5.4251$ N/m and $\sigma^s = 0.5689$ N/m (Tian, 2006). The radius of the inclusion is $R_{inclusion} = 1$ nm.

A series of parametric studies on interface properties are conducted to investigate the influence of surface bending resistance on stress distributions in vicinity of the inclusion. Fig. 6 shows variations of stress components along the positive $y-\text{axis}$ with different interface properties and different far-field loading. As seen from the results, stress distributions in the inclusion with the Gurtin-Murdoch model differ a lot from the results with the Steigmann–Ogden model. Moreover, the stress component with different interface bending stiffness parameters have the same value along the circle $R=R_{st}, \theta=\frac{\pi}{2}$ (the value of $R_{st}$ is not presented in this paper because the expression for $R_{st}$ is too length) in the inclusion, which is similar to the case of a nano-void.

We also study the influence of interface stiffness parameters on the stress components at the point $(0,0,0)$ in the inclusion, as shown in Fig.7. It is easily observed that as $(\chi^s, \zeta^s)$ gets close to the characteristic line:

$$5\chi^s + 3\zeta^s + c_{inclusion} = 0 \tag{37}$$

the value of $\sigma_{zz}^i$ will become extremely large. The constant $c_{inclusion}$ is given in Appendix A. This reveals that the interface bending stiffness parameters can significantly change the stress distributions in the inclusion when $(\chi^s, \zeta^s)$ gets close to the characteristic line.

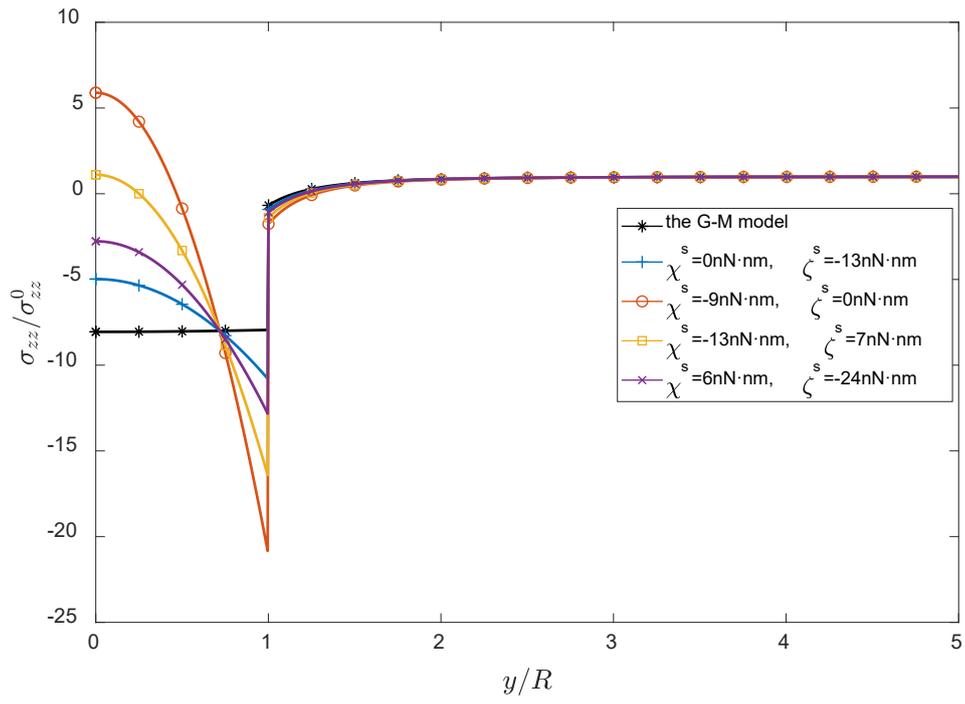

(a)

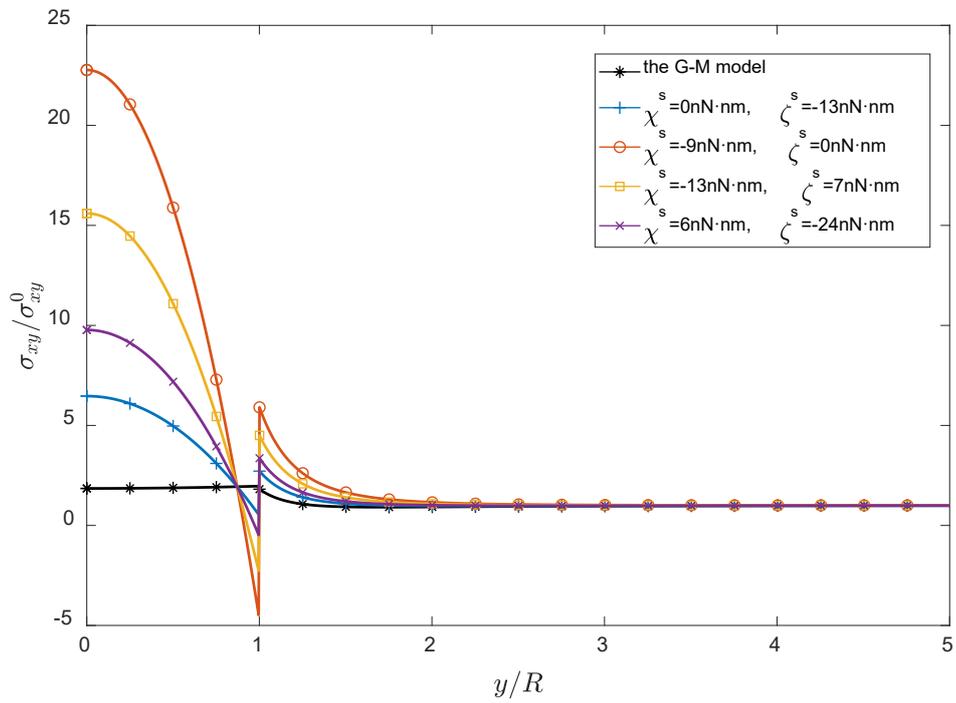

(b)

Fig. 6. (a)Variation of $\sigma_{zz}$ along the positive $y$–axis for different interface

properties when $\boldsymbol{\sigma}^0 = \begin{pmatrix} 0 & 0 & 0 \\ 0 & 0 & 0 \\ 0 & 0 & 100 \end{pmatrix}$ MPa and (b) variation of $\sigma_{xy}$ along the positive

$y-$axis for different interface properties when $\boldsymbol{\sigma}^0 = \begin{pmatrix} 0 & 100 & 0 \\ 100 & 0 & 0 \\ 0 & 0 & 0 \end{pmatrix}$ MPa.

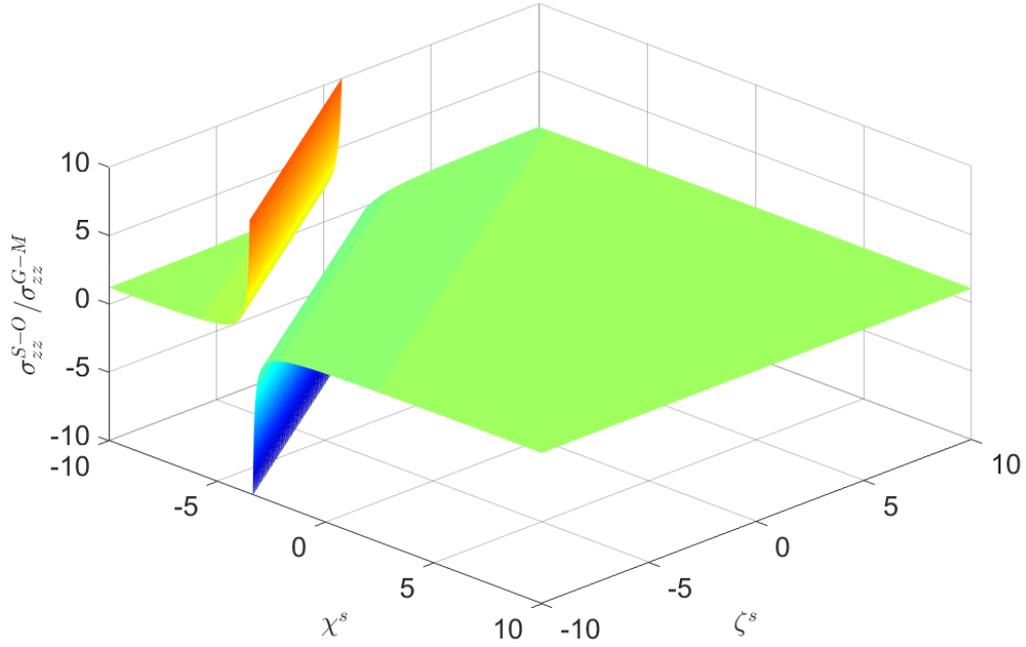

Fig. 7. The influence of $\chi^s$ and $\zeta^s$ on $\sigma_{zz}^i$ at the point $(0,0,0)$ when

$\boldsymbol{\sigma}^0 = \begin{pmatrix} 0 & 0 & 0 \\ 0 & 0 & 0 \\ 0 & 0 & 100 \end{pmatrix}$ MPa is applied. The stress from the Steigmann–Ogden interface model

$\sigma_{zz}^{\text{S-O}}$ is normalized by that from the Gurtin-Murdoch interface model $\sigma_{zz}^{\text{G-M}}$

Fig. 8. shows computed $\sigma_{zz}^i / \sigma_{zz}^0$ at point $(0, R_{\text{inclusion}}, 0)$ with different interface stiffness parameters and with different inclusion radiuses. The results reveal that the interface effect caused by interface bending stiffness parameters is size dependent. The smaller the inclusion is, the more significant interface effects are.

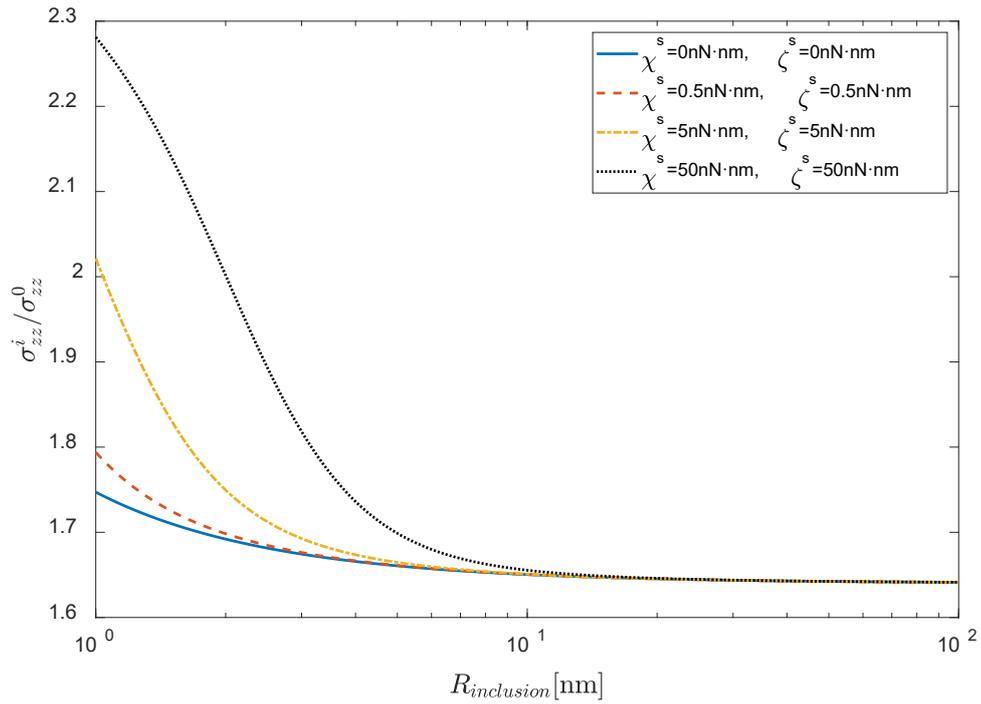

Fig. 8. $\sigma_{rr}^i / \sigma^s$ at point $(0, R_{inclusion}, 0)$ in the inclusion with different inclusion radiuses

## 6. Conclusions

In this study, an explicit solution considering the interface bending resistance based on the Steigmann–Ogden interface model is derived for the first time, for a spherical nano-inhomogeneity (nanoscale void/inclusion) embedded in an infinite matrix under general uniform far-field-stress (including both tensions and shears). Numerical examples show that the stress fields considering the interface bending resistance with the Steigmann–Ogden interface model, differ significantly from those considering only the interface stretching resistance with the Gurtin–Murdoch interface model.

Two interesting phenomena are observed in this study. First, we observe that some stress components are invariant to interface bending stiffness parameters along a certain circle in the inclusion/matrix when interface stretch stiffness parameters are fixed. Secondly, we presented a characteristic line for interface bending stiffness parameters in this paper. If the interface bending stiffness parameters get close to the characteristic line, the stress concentration phenomenon will become quite severe, which should be carefully considered when designing nanocomposites and porous materials.

The explicit analytical solution derived in this paper can be used as a benchmark for validating numerical methods for modelling composites and porous materials, such as FEM as well as the method of computational grains (Dong and Atluri, 2012a, b; Wang et al., 2018c) which is currently being developed by the authors. The derived analytical solution with the Steigmann–Ogden interface model is provided in the supplemental MATLAB code for the convenience of users.

**Acknowledgement**

The authors thankfully acknowledge the support from the National Key Research and Development Program of China (No. 2017YFA0207800) and the National Natural Science Foundation of China (grant No. 11872008).

**Appendix A**

For the case that the remote loading has only one non-zero stress component being $\sigma_{kk}^0 (k=x,y,z)$, The dimensionless constants $M_{pkk} (p=1,...,5 \text{ and } k=x,y,z)$ and $C_{qkk} (q=1,2,3 \text{ and } k=x,y,z)$ are defined by:

$$M_{1kk} = -\frac{(-1+2v^m)\sigma_{kk}^0}{6(1+v^m)\mu^m} \tag{A.1}$$

$$\begin{aligned}M_{2kk} = (R^3((-1+2v^m)((-2+4v^i)\lambda^s - R(1+v^i)\mu^i + (-1+2v^i)(2\mu^s + \\ \sigma^s))\sigma_{kk}^0 - (-1+2v^i)(1+v^m)\mu^m(6\sigma^s - R\sigma_{kk}^0)))/(6(1+v^m)\mu^m((-2+4v^i \\ )\lambda^s - R(\mu^i + v^i\mu^i + 2\mu^m - 4v^i\mu^m) + (-1+2v^i)(2\mu^s + \sigma^s)))\end{aligned} \tag{A.2}$$

$$M_{3kk} = \frac{\sigma_{kk}^0}{6\mu^m} \tag{A.3}$$

$$\begin{aligned}M_{4kk} = -((5R^3\sigma_{kk}^0(R^4(\mu^i - \mu^m)((7+5v^i)\mu^i + 4(7-10v^i)\mu^m) - 3\zeta^s(8 \\ (-7+10v^i)\lambda^s + R(-49\mu^i + 61v^i\mu^i + 28\mu^m - 40v^i\mu^m) + 2(-7+10v^i)( \\ 10\mu^s - \sigma^s)) + R^3((35-47v^i)\lambda^s\mu^i + 4(-7+10v^i)\lambda^s\mu^m - 49(-1+v^i)\mu^i \\ \mu^s + (35-53v^i)\mu^i\sigma^s + 6(-7+10v^i)\mu^m\sigma^s) - 2R^2(-7+10v^i)(2\mu^s(\lambda^s + \\ \mu^s) + (\lambda^s + 5\mu^s)\sigma^s - \sigma^{s2}) - 5R((-49+61v^i)\mu^i + 4(7-10v^i)\mu^m)\chi^s - \\ 10(-7+10v^i)(4\lambda^s + 10\mu^s - \sigma^s)\chi^s))/(12\mu^m(-R^4((7+5v^i)\mu^i + 4(7- \\ 10v^i)\mu^m)(2(-4+5v^m)\mu^i + (-7+5v^m)\mu^m) + 6\zeta^s(8(-7+10v^i)(-4+5v^m) \\ \lambda^s + R(-49+61v^i)(-4+5v^m)\mu^i + 4R(-7+10v^i)(-8+7v^m)\mu^m + 2(-7 \\ +10v^i)(-4+5v^m)(10\mu^s - \sigma^s)) + 4R^2(-7+10v^i)(-4+5v^m)(2\mu^s(\lambda^s + \mu^s \\ ) + (\lambda^s + 5\mu^s)\sigma^s - \sigma^{s2}) + 2R^3((-35+47v^i)(-4+5v^m)\lambda^s\mu^i + 4(-7+10v^i \\ )(-5+4v^m)\lambda^s\mu^m + (-4+5v^m)\mu^i(49(-1+v^i)\mu^s + (-35+53v^i)\sigma^s) + 3(-7 \\ +10v^i)\mu^m(14(-1+v^m)\mu^s + (-5+3v^m)\sigma^s)) + 10R((-49+61v^i)(-4+5v^m \\ )\mu^i + 4(-7+10v^i)(-8+7v^m)\mu^m)\chi^s + 20(-7+10v^i)(-4+5v^m)(4\lambda^s + 10 \\ \mu^s - \sigma^s)\chi^s)))\end{aligned} \tag{A.4}$$

$$\begin{aligned}M_{5kk} = -((R^5\sigma_{kk}^0(R^4(\mu^i - \mu^m)((7+5v^i)\mu^i + 4(7-10v^i)\mu^m) - 3\zeta^s(8( \\ -7+10v^i)\lambda^s + R(-49+61v^i)\mu^i + 4R(-7+10v^i)(-3+2v^m)\mu^m + 2(-7 \\ +10v^i)(10\mu^s - \sigma^s)) - 2R^2(-7+10v^i)(2\mu^s(\lambda^s + \mu^s) + (\lambda^s + 5\mu^s)\sigma^s - \\ \sigma^{s2}) + R^3(\lambda^s((35-47v^i)\mu^i + 4(-7+10v^i)v^m\mu^m) + \mu^i(-49(-1+v^i)\mu^s \\ +35\sigma^s - 53v^i\sigma^s) + 2(-7+10v^i)\mu^m(4(-1+v^m)\mu^s + 5\sigma^s - 2v^m\sigma^s)) - 5R \\ ((-49+61v^i)\mu^i + 4(-7+10v^i)(-3+2v^m)\mu^m)\chi^s - 10(-7+10v^i)(4\lambda^s \\ +10\mu^s - \sigma^s)\chi^s))/(2\mu^m(-R^4((7+5v^i)\mu^i + 4(7-10v^i)\mu^m)(2(-4+5v^m \\ )\mu^i + (-7+5v^m)\mu^m) + 6\zeta^s(8(-7+10v^i)(-4+5v^m)\lambda^s + R(-49+61v^i)( \\ -4+5v^m)\mu^i + 4R(-7+10v^i)(-8+7v^m)\mu^m + 2(-7+10v^i)(-4+5v^m)(10 \\ \mu^s - \sigma^s)) + 4R^2(-7+10v^i)(-4+5v^m)(2\mu^s(\lambda^s + \mu^s) + (\lambda^s + 5\mu^s)\sigma^s - \\ \sigma^{s2}) + 2R^3((-35+47v^i)(-4+5v^m)\lambda^s\mu^i + 4(-7+10v^i)(-5+4v^m)\lambda^s\mu^m \\ +(-4+5v^m)\mu^i(49(-1+v^i)\mu^s + (-35+53v^i)\sigma^s) + 3(-7+10v^i)\mu^m(14( \\ -1+v^m)\mu^s + (-5+3v^m)\sigma^s)) + 10R((-49+61v^i)(-4+5v^m)\mu^i + 4(-7+ \\ 10v^i)(-8+7v^m)\mu^m)\chi^s + 20(-7+10v^i)(-4+5v^m)(4\lambda^s + 10\mu^s - \sigma^s)\chi^s)))\end{aligned} \tag{A.5}$$

$$C_{1kk} = -(((-1+2v^i)(2(1+v^m)\sigma^s + R(-1+v^m)\sigma^0_{kk}))/(2(1+v^m)((-2+ \\ 4v^i)\lambda^s - R(\mu^i + v^i\mu^i + 2\mu^m - 4v^i\mu^m) + (-1+2v^i)(2\mu^s + \sigma^s))))$$ (A.6)

$$C_{2kk} = (5(-1+v^m)\sigma^0_{kk}(6\zeta^s + R^2(-\lambda^s - 2\mu^s + \sigma^s) + 10\chi^s))/(R(R^4((7 \\ +5v^i)\mu^i + 4(7-10v^i)\mu^m)(2(-4+5v^m)\mu^i + (-7+5v^m)\mu^m) - 6\zeta^s(8(-7 \\ +10v^i)(-4+5v^m)\lambda^s + R(-49+61v^i)(-4+5v^m)\mu^i + 4R(-7+10v^i)(-8 \\ +7v^m)\mu^m + 2(-7+10v^i)(-4+5v^m)(10\mu^s - \sigma^s)) - 4R^2(-7+10v^i)(-4 \\ +5v^m)(2\mu^s(\lambda^s + \mu^s) + (\lambda^s + 5\mu^s)\sigma^s - \sigma^{s2}) - 2R^3((-35+47v^i)(-4+5 \\ v^m)\lambda^s\mu^i + 4(-7+10v^i)(-5+4v^m)\lambda^s\mu^m + (-4+5v^m)\mu^i(49(-1+v^i)\mu^s \\ +(-35+53v^i)\sigma^s) + 3(-7+10v^i)\mu^m(14(-1+v^m)\mu^s + (-5+3v^m)\sigma^s)) - \\ 10R((-49+61v^i)(-4+5v^m)\mu^i + 4(-7+10v^i)(-8+7v^m)\mu^m)\chi^s - 20( \\ -7+10v^i)(-4+5v^m)(4\lambda^s + 10\mu^s - \sigma^s)\chi^s))$$ (A.7)

$$C_{3kk} = -((5R(-1+v^m)\sigma^0_{kk}(12(-7+16v^i)\zeta^s - R^3((7+5v^i)\mu^i + 4(7-10 \\ v^i)\mu^m) + R^2(6(-7+8v^i)\lambda^s + 56(-1+v^i)\mu^s + 4(-7+13v^i)\sigma^s) + 20(-7 \\ +16v^i)\chi^s))/(2(R^4((7+5v^i)\mu^i + 4(7-10v^i)\mu^m)(2(-4+5v^m)\mu^i + (-7 \\ +5v^m)\mu^m) - 6\zeta^s(8(-7+10v^i)(-4+5v^m)\lambda^s + R(-49+61v^i)(-4+5v^m) \\ \mu^i + 4R(-7+10v^i)(-8+7v^m)\mu^m + 2(-7+10v^i)(-4+5v^m)(10\mu^s - \sigma^s) \\ ) - 4R^2(-7+10v^i)(-4+5v^m)(2\mu^s(\lambda^s + \mu^s) + (\lambda^s + 5\mu^s)\sigma^s - \sigma^{s2}) - 2R^3 \\ ((-35+47v^i)(-4+5v^m)\lambda^s\mu^i + 4(-7+10v^i)(-5+4v^m)\lambda^s\mu^m + (-4+5v^m \\ )\mu^i(49(-1+v^i)\mu^s + (-35+53v^i)\sigma^s) + 3(-7+10v^i)\mu^m(14(-1+v^m)\mu^s + \\ (-5+3v^m)\sigma^s)) - 10R((-49+61v^i)(-4+5v^m)\mu^i + 4(-7+10v^i)(-8+7v^m \\ )\mu^m)\chi^s - 20(-7+10v^i)(-4+5v^m)(4\lambda^s + 10\mu^s - \sigma^s)\chi^s)))$$ (A.8)

For the case that the remote loading has only one non-zero stress component being $\sigma^0_{st}(s,t = x,y,z, s \neq t)$, The dimensionless constants $M_{pst}(p=2,...,5 \text{ and } s,t=x,y,z)$ and $C_{qst}(q=1,2,3 \text{ and } s,t=x,y,z)$ are defined by:

$$M_{2st} = (R^3(-1+2v^i)\sigma^s)/((2-4v^i)\lambda^s + R(\mu^i + v^i\mu^i + 2\mu^m - 4v^i\mu^m) \\ +2\mu^s + \sigma^s - 2v^i(2\mu^s + \sigma^s))$$ (A.9)

$$M_{3st} = \frac{\sigma^0_{st}}{6\mu^m};$$ (A.10)

$$M_{4st} = -((5R^3\sigma_{st}^0(R^4(\mu^i - \mu^m)((7+5v^i)\mu^i + 4(7-10v^i)\mu^m) - 3\zeta^s(8($$
$$-7+10v^i)\lambda^s + R(-49\mu^i + 61v^i\mu^i + 28\mu^m - 40v^i\mu^m) + 2(-7+10v^i)(10$$
$$\mu^s - \sigma^s)) + R^3((35-47v^i)\lambda^s\mu^i + 4(-7+10v^i)\lambda^s\mu^m - 49(-1+v^i)\mu^i\mu^s$$
$$+(35-53v^i)\mu^i\sigma^s + 6(-7+10v^i)\mu^m\sigma^s) - 2R^2(-7+10v^i)(2\mu^s(\lambda^s + \mu^s)$$
$$+(\lambda^s + 5\mu^s)\sigma^s - \sigma^{s2}) - 5R((-49+61v^i)\mu^i + 4(7-10v^i)\mu^m)\chi^s - 10(-$$
$$7+10v^i)(4\lambda^s + 10\mu^s - \sigma^s)\chi^s))/(12\mu^m(-R^4((7+5v^i)\mu^i + 4(7-10v^i)$$
$$\mu^m)(2(-4+5v^m)\mu^i + (-7+5v^m)\mu^m) + 6\zeta^s(8(-7+10v^i)(-4+5v^m)\lambda^s +$$
$$R(-49+61v^i)(-4+5v^m)\mu^i + 4R(-7+10v^i)(-8+7v^m)\mu^m + 2(-7+10v^i$$
$$)(-4+5v^m)(10\mu^s - \sigma^s)) + 4R^2(-7+10v^i)(-4+5v^m)(2\mu^s(\lambda^s + \mu^s) + ($$
$$\lambda^s + 5\mu^s)\sigma^s - \sigma^{s2}) + 2R^3((-35+47v^i)(-4+5v^m)\lambda^s\mu^i + 4(-7+10v^i)$$
$$(-5+4v^m)\lambda^s\mu^m + (-4+5v^m)\mu^i(49(-1+v^i)\mu^s + (-35+53v^i)\sigma^s) + 3($$
$$-7+10v^i)\mu^m(14(-1+v^m)\mu^s + (-5+3v^m)\sigma^s)) + 10R((-49+61v^i)(-4$$
$$+5v^m)\mu^i + 4(-7+10v^i)(-8+7v^m)\mu^m)\chi^s + 20(-7+10v^i)(-4+5v^m)(4$$
$$\lambda^s + 10\mu^s - \sigma^s)\chi^s)))$$

(A.11)

$$M_{5st} = -((R^5\sigma_{st}^0(R^4(\mu^i - \mu^m)((7+5v^i)\mu^i + 4(7-10v^i)\mu^m) - 3\zeta^s(8(-$$
$$7+10v^i)\lambda^s + R(-49+61v^i)\mu^i + 4(-7+10v^i)(-3+2v^m)\mu^m + 2(-7$$
$$+10v^i)(10\mu^s - \sigma^s)) - 2R^2(-7+10v^i)(2\mu^s(\lambda^s + \mu^s) + (\lambda^s + 5\mu^s)\sigma^s -$$
$$\sigma^{s2}) + R^3(\lambda^s((35-47v^i)\mu^i + 4(-7+10v^i)v^m\mu^m) + \mu^i(-49(-1+v^i)\mu^s$$
$$+35\sigma^s - 53v^i\sigma^s) + 2(-7+10v^i)\mu^m(4(-1+v^m)\mu^s + 5\sigma^s - 2v^m\sigma^s)) - 5$$
$$R((-49+61v^i)\mu^i + 4(-7+10v^i)(-3+2v^m)\mu^m)\chi^s - 10(-7+10v^i)(4\lambda^s$$
$$+10\mu^s - \sigma^s)\chi^s))/(2\mu^m(-R^4((7+5v^i)\mu^i + 4(7-10v^i)\mu^m)(2(-4+5v^m$$
$$)\mu^i + (-7+5v^m)\mu^m) + 6\zeta^s(8(-7+10v^i)(-4+5v^m)\lambda^s + R(-49+61v^i)($$
$$-4+5v^m)\mu^i + 4R(-7+10v^i)(-8+7v^m)\mu^m + 2(-7+10v^i)(-4+5v^m)(10$$
$$\mu^s - \sigma^s)) + 4R^2(-7+10v^i)(-4+5v^m)(2\mu^s(\lambda^s + \mu^s) + (\lambda^s + 5\mu^s)\sigma^s -$$
$$\sigma^{s2}) + 2R^3((-35+47v^i)(-4+5v^m)\lambda^s\mu^i + 4(-7+10v^i)(-5+4v^m)\lambda^s\mu^m$$
$$+(-4+5v^m)\mu^i(49(-1+v^i)\mu^s + (-35+53v^i)\sigma^s) + 3(-7+10v^i)\mu^m(14($$
$$-1+v^m)\mu^s + (-5+3v^m)\sigma^s)) + 10R((-49+61v^i)(-4+5v^m)\mu^i + 4(-7+$$
$$10v^i)(-8+7v^m)\mu^m)\chi^s + 20(-7+10v^i)(-4+5v^m)(4\lambda^s + 10\mu^s - \sigma^s)\chi^s)))$$

(A.12)

$$C_{1st} = (\sigma^s - 2v^i\sigma^s)/((-2+4v^i)\lambda^s - R(\mu^i + v^i\mu^i + 2\mu^m - 4v^i\mu^m) + (-1$$
$$+2v^i)(2\mu^s + \sigma^s))$$

(A.13)

$$C_{2st} = (5(-1+v^m)\sigma_{st}^0(6\zeta^s + R^2(-\lambda^s - 2\mu^s + \sigma^s) + 10\chi^s))/(R(R^4((7+5v^i)\mu^i + 4(7-10v^i)\mu^m)(2(-4+5v^m)\mu^i + (-7+5v^m)\mu^m) - 6\zeta^s(8(-7+10v^i)(-4+5v^m)\lambda^s + R(-49+61v^i)(-4+5v^m)\mu^i + 4R(-7+10v^i)(-8+7v^m)\mu^m + 2(-7+10v^i)(-4+5v^m)(10\mu^s - \sigma^s)) - 4R^2(-7+10v^i)(-4+5v^m)(2\mu^s(\lambda^s + \mu^s) + (\lambda^s + 5\mu^s)\sigma^s - \sigma^{s2}) - 2R^3((-35+47v^i)(-4+5v^m)\lambda^s\mu^i + 4(-7+10v^i)(-5+4v^m)\lambda^s\mu^m + (-4+5v^m)\mu^i(49(-1+v^i)\mu^s + (-35+53v^i)\sigma^s) + 3(-7+10v^i)\mu^m(14(-1+v^m)\mu^s + (-5+3v^m)\sigma^s)) - 10R((-49+61v^i)(-4+5v^m)\mu^i + 4(-7+10v^i)(-8+7v^m)\mu^m)\chi^s - 20(-7+10v^i)(-4+5v^m)(4\lambda^s + 10\mu^s - \sigma^s)\chi^s))$$
(A.14)

$$C_{3st} = -((5R(-1+v^m)\sigma_{st}^0(12(-7+16v^i)\zeta^s - R^3((7+5v^i)\mu^i + 4(7-10v^i)\mu^m) + R^2(6(-7+8v^i)\lambda^s + 56(-1+v^i)\mu^s + 4(-7+13v^i)\sigma^s) + 20(-7+16v^i)\chi^s))/(2(R^4((7+5v^i)\mu^i + 4(7-10v^i)\mu^m)(2(-4+5v^m)\mu^i + (-7+5v^m)\mu^m) - 6\zeta^s(8(-7+10v^i)(-4+5v^m)\lambda^s + R(-49+61v^i)(-4+5v^m)\mu^i + 4R(-7+10v^i)(-8+7v^m)\mu^m + 2(-7+10v^i)(-4+5v^m)(10\mu^s - \sigma^s)) - 4R^2(-7+10v^i)(-4+5v^m)(2\mu^s(\lambda^s + \mu^s) + (\lambda^s + 5\mu^s)\sigma^s - \sigma^{s2}) - 2R^3((-35+47v^i)(-4+5v^m)\lambda^s\mu^i + 4(-7+10v^i)(-5+4v^m)\lambda^s\mu^m + (-4+5v^m)\mu^i(49(-1+v^i)\mu^s + (-35+53v^i)\sigma^s) + 3(-7+10v^i)\mu^m(14(-1+v^m)\mu^s + (-5+3v^m)\sigma^s)) - 10R((-49+61v^i)(-4+5v^m)\mu^i + 4(-7+10v^i)(-8+7v^m)\mu^m)\chi^s - 20(-7+10v^i)(-4+5v^m)(4\lambda^s + 10\mu^s - \sigma^s)\chi^s)))$$
(A.15)

For void problem, we can get $M_{pst}$ by setting $v_i = \mu_i = 0$ in Eqs. (A.1- A.5 and A.9-A.12).

Other constants in this paper is given here:

$$c_{void} = (R^2(2R^2(-7+5v^m)\mu^{m2} + R\mu^m(4(-5+4v^m)\lambda s + 42(-1+v^m)\mu s + 3(-5+3v^m)\sigma^s) + 2(-4+5v^m)(2\mu^s(\lambda s + \mu^s) + (\lambda s + 5\mu^s)\sigma^s - \sigma^{s2}))) / (8(-4+5v^m)\lambda s + 4R(-8+7v^m)\mu^m + 2(-4+5v^m)(10\mu^s - \sigma^s))$$
(A.16)

$$c_{inclusion} = (R^2(-R^2((7+5v^i)\mu^i + 4(7-10v^i)\mu^m)(2(-4+5v^m)\mu^i + (-7+5v^m)\mu^m) + 4(-7+10v^i)(-4+5v^m)(2\mu^s(\lambda^s + \mu^s) + (\lambda^s + 5\mu^s)\sigma^s - \sigma^{s2}) + 2R((-35+47v^i)(-4+5v^m)\lambda^s\mu^i + 4(-7+10v^i)(-5+4v^m)\lambda^s\mu^m + (-4+5v^m)\mu^i(49(-1+v^i)\mu^s + (-35+53v^i)\sigma^s) + 3(-7+10v^i)\mu^m(14(-1+v^m)\mu^s + (-5+3v^m)\sigma^s))))/(2(8(-7+10v^i)(-4+5v^m)\lambda^s + R(-49+61v^i)(-4+5v^m)\mu^i + 4R(-7+10v^i)(-8+7v^m)\mu^m + 2(-7+10v^i)(-4+5v^m)(10\mu^s - \sigma^s)))$$
(A.17)

$$\begin{aligned}
\theta_{void} = \frac{1}{2}\arctan[&-((\sqrt{(((8(88+(29-167v^m)v^m)\lambda^{s2}-1724R^2\mu^{m2}+3188} \\
&R^2 v^m\mu^{m2} -1424R^2 v^{m2}\mu^{m2} -7004R\mu^m\mu^s +19568Rv^m\mu^m\mu^s -11444R \\
&v^{m2}\mu^m\mu^s -6936\mu^{s2} +27132v^m\mu^{s2} -22956v^{m2}\mu^{s2} +3286R\mu^m\sigma^s -6172 \\
&Rv^m\mu^m\sigma^s +2926Rv^{m2}\mu^m\sigma^s +7208\mu^s\sigma^s -18296v^m\mu^s\sigma^s +11648v^{m2}\mu^s\sigma^s \\
&-1166\sigma^{s2} +2597v^m\sigma^{s2} -1421v^{m2}\sigma^{s2} +2\lambda^s(-6R(18+v^m(-301+233v^m) \\
&)\mu^m +2(68+(2209-2827v^m)v^m)\mu^s +3(212+v^m(-659+497v^m))\sigma^s)- \\
&32\lambda^s \sqrt{(5(6(2+(2(-2+v^m)\lambda^s +2R(-8+77v^m)\mu^m -34\mu^s +56v^m\mu^s +} \\
&13\sigma^s -14v^m\sigma^s)+((-32+22v^m)\lambda^s +2R(-13+8v^m)\mu^m -68\mu^s +58v^m \\
&\mu^s +2\sigma^s -7v^m\sigma^s)^2) +22v^m\lambda^s \sqrt{(5(6(2+3v^m)\lambda^s -20R\mu^m +22Rv^m\mu^m} \\
&-34\mu^s +92v^m\mu^s +29\sigma^s -28v^m\sigma^s)(2(-2+7v^m)\lambda^s +2R(-8+7v^m)\mu^m \\
&-34\mu^s +56v^m\mu^s +13\sigma^s -14v^m\sigma^s)+((-32+22v^m)\lambda^s +2R(-13+8v^m) \\
&\mu^m -68\mu^s +58v^m\mu^s +2\sigma^s -7v^m\sigma^s)^2) -26R\mu^m \sqrt{(5(6(2+3v^m)\lambda^s -20} \\
&R\mu^m +22Rv^m\mu^m -34\mu^s +92v^m\mu^s +29\sigma^s -28v^m\sigma^s)(2(-2+7v^m)\lambda^s + \\
&2R(-8+7v^m)\mu^m -34\mu^s +56v^m\mu^s +13\sigma^s -14v^m\sigma^s)+((-32+22v^m) \\
&\lambda^s +2R(-13+8v^m)\mu^m -68\mu^s +58v^m\mu^s +2\sigma^s -7v^m\sigma^s)^2) +16Rv^m\mu^m \\
&\sqrt{(5(6(2+3v^m)\lambda^s -20R\mu^m +22Rv^m\mu^m -34\mu^s +92v^m\mu^s +29\sigma^s -28v^m} \\
&\sigma^s)(2(-2+7v^m)\lambda^s +2R(-8+7v^m)\mu^m -34\mu^s +56v^m\mu^s +13\sigma^s -14v^m \\
&\sigma^s)+((-32+22v^m)\lambda^s +2R(-13+8v^m)\mu^m -68\mu^s +58v^m\mu^s +2\sigma^s -7 \\
&v^m\sigma^s)^2) -68\mu^s \sqrt{(5(6(2+3v^m)\lambda^s -20R\mu^m +22Rv^m\mu^m -34\mu^s +92v^m} \\
&\mu^s +29\sigma^s -28v^m\sigma^s)(2(-2+7v^m)\lambda^s +2R(-8+7v^m)\mu^m -34\mu^s +56v^m \\
&\mu^s +13\sigma^s -14v^m\sigma^s)+((-32+22v^m)\lambda^s +2R(-13+8v^m)\mu^m -68\mu^s + \\
&58v^m\mu^s +2\sigma^s -7v^m\sigma^s)^2) +58v^m\mu^s \sqrt{(5(6(2+3v^m)\lambda^s -20R\mu^m +22R} \\
&v^m\mu^m -34\mu^s +92v^m\mu^s +29\sigma^s -28v^m\sigma^s)(2(-2+7v^m)\lambda^s +2R(-8+7 \\
&v^m)\mu^m -34\mu^s +56v^m\mu^s +13\sigma^s -14v^m\sigma^s)+((-32+22v^m)\lambda^s +2R(- \\
&13+8v^m)\mu^m -68\mu^s +58v^m\mu^s +2\sigma^s -7v^m\sigma^s)^2) +2\sigma^s \sqrt{(5(6(2+3v^m)\lambda^s} \\
&-20R\mu^m +22Rv^m\mu^m -34\mu^s +92v^m\mu^s +29\sigma^s -28v^m\sigma^s)(2(-2+7v^m)\lambda^s \\
&+2R(-8+7v^m)\mu^m -34\mu^s +56v^m\mu^s +13\sigma^s -14v^m\sigma^s)+((-32+22v^m) \\
&\lambda^s +2R(-13+8v^m)\mu^m -68\mu^s +58v^m\mu^s +2\sigma^s -7v^m\sigma^s)^2) -7v^m\sigma^s \sqrt{(5} \\
&(6(2+3v^m)\lambda^s -20R\mu^m +22Rv^m\mu^m -34\mu^s +92v^m\mu^s +29\sigma^s -28v^m\sigma^s) \\
&(2(-2+7v^m)\lambda^s +2R(-8+7v^m)\mu^m -34\mu^s +56v^m\mu^s +13\sigma^s -14v^m\sigma^s)+ \\
&((-32+22v^m)\lambda^s +2R(-13+8v^m)\mu^m -68\mu^s +58v^m\mu^s +2\sigma^s -7v^m\sigma^s)^2 \\
&))/(2(-2+7v^m)\lambda^s +2R(-8+7v^m)\mu^m -34\mu^s +56v^m\mu^s +13\sigma^s -14v^m \\
&\sigma^s)))/(\sqrt{((2-7v^m)\lambda^s +R(8-7v^m)\mu^m +17\mu^s -13\sigma^s/2+7v^m(-4\mu^s +} \\
&))/(-((-32\lambda^s +22v^m\lambda^s -26R\mu^m +16Rv^m\mu^m -68\mu^s +58v^m\mu^s +2\sigma^s - \\
&7v^m\sigma^s +\sqrt{(5(6(2+3v^m)\lambda^s -20R\mu^m +22Rv^m\mu^m -34\mu^s +92v^m\mu^s +29\sigma^s} \\
&-28v^m\sigma^s)(2(-2+7v^m)\lambda^s +2R(-8+7v^m)\mu^m -34\mu^s +56v^m\mu^s +13\sigma^s - \\
&14v^m\sigma^s)+((-32+22v^m)\lambda^s +2R(-13+8v^m)\mu^m -68\mu^s +58v^m\mu^s +2\sigma^s \\
&-7v^m\sigma^s)^2))/(2(-2+7v^m)\lambda^s +2R(-8+7v^m)\mu^m -34\mu^s +56v^m\mu^s +13\sigma^s \\
&-14v^m\sigma^s)))]
\end{aligned}$$

(A.18)

## Appendix B

In Appendix B, a brief introduction to the derivation process is given.

Firstly, we give a brief introduction to the Papkovich-Neuber potentials used in this paper. Substitute Eq.(24) into Eq.(20), and we can obtain the positive $n-th$ $(n \geq 0)$ order Papkovich-Neuber solutions. For example, the $0^{th}$ order Papkovich-Neuber solutions can be written as:

$$\boldsymbol{\alpha}_0 = \frac{1}{\mu^j \sqrt{\pi}} \begin{pmatrix} 1-v^j & 0 & 0 \\ 0 & 1-v^j & 0 \\ 0 & 0 & 1-v^j \end{pmatrix} \tag{B.1}$$

And the 1st order Papkovich-Neuber solutions can be written as:

$$\boldsymbol{\alpha}_1 = \frac{\sqrt{\frac{3}{2\pi}}}{8\mu^j} \begin{pmatrix} r(-5+4v^j)\cos[\theta] & 0 & r(-5+4v^j)\sin[\theta]\sin[\varphi] \\ 0 & r(-5+4v^j)\cos[\theta] & r(5-4v^j)\cos[\varphi]\sin[\theta] \\ r(5-4v^j)\cos[\varphi]\sin[\theta] & r(5-4v^j)\sin[\theta]\sin[\varphi] & 0 \end{pmatrix}$$

$$\begin{matrix} -2\sqrt{2}r\cos[\varphi]\sin[\theta] & 2r\cos[\varphi]\sin[\theta] & 0 \\ -2\sqrt{2}r\sin[\theta]\sin[\varphi] & 8r(-1+v^j)\sin[\theta]\sin[\varphi] & r(-5+4v^j)\cos[\theta] \\ -8\sqrt{2}r(-1+v^j)\cos[\theta] & 2r\cos[\theta] & r(-5+4v^j)\sin[\theta]\sin[\varphi] \\ 8r(-1+v^j)\cos[\varphi]\sin[\theta] & r(-5+4v^j)\sin[\theta]\sin[\varphi] & r(-5+4v^j)\cos[\theta] \\ 2r\sin[\theta]\sin[\varphi] & r(-5+4v^j)\cos[\varphi]\sin[\theta] & 0 \\ 2r\cos[\theta] & 0 & r(-5+4v^j)\cos[\varphi]\sin[\theta] \end{matrix}$$

(B.2)

It should be pointed out that n-th order Papkovich-Neuber solutions is written as a $3 \times (6n+3)$ matrix, while the 3 rows represent the 3 displacement solutions, and the $6n+3$ columns represent independent modes produced by different harmonics.

Substitute Eq.(25) into Eq.(19), and we can obtain the negative $n-th$ order Papkovich-Neuber solutions. For example, the -1st order Papkovich-Neuber potentials can be written as:

$$\boldsymbol{\alpha}_{-1} = \frac{1}{4\mu^j \sqrt{\pi r}} \begin{pmatrix} 4-4v^j - \cos[\theta]^2 \cos[\varphi]^2 - \sin[\varphi]^2 & \frac{1}{2}\sin[\theta]^2 \sin[2\varphi] & \cos[\theta]\cos[\varphi]\sin[\theta] \\ \frac{1}{2}\sin[\theta]^2 \sin[2\varphi] & 4-4v^j - \cos[\varphi]^2 - \cos[\theta]^2 \sin[\varphi]^2 & \cos[\theta]\sin[\theta]\sin[\varphi] \\ \cos[\theta]\cos[\varphi]\sin[\theta] & \cos[\theta]\sin[\theta]\sin[\varphi] & \frac{1}{2}(7-8v^j + \cos[2\theta]) \end{pmatrix}$$
(B.3)

The displacement field in the matrix/inclusion can then be expressed as a linear combination of the obtained Papkovich-Neuber solutions:

$$\begin{aligned} \mathbf{u}^m &= \begin{bmatrix} \boldsymbol{\alpha}_{-p} & \boldsymbol{\alpha}_{-p+1} & \cdots & \boldsymbol{\alpha}_{q-1} & \boldsymbol{\alpha}_q \end{bmatrix} \mathbf{x} \\ \mathbf{u}^i &= \begin{bmatrix} \boldsymbol{\alpha}_0 & \boldsymbol{\alpha}_1 & \cdots & \boldsymbol{\alpha}_{t-1} & \boldsymbol{\alpha}_t \end{bmatrix} \mathbf{y} \end{aligned}$$
(B.4)

where $\mathbf{x}$ and $\mathbf{y}$ are the to be determined coefficient vector. For the current problem, only $\boldsymbol{\alpha}_{-4}$, $\boldsymbol{\alpha}_{-2}$, $\boldsymbol{\alpha}_0$, $\boldsymbol{\alpha}_1$ are needed for the matrix, and $\boldsymbol{\alpha}_1$, $\boldsymbol{\alpha}_3$ are needed for the inhomogeneity. Using the strain displacement-gradient compatibility and the constitutive relations, the stress fields can be obtained easily. The stress field correspond to $\boldsymbol{\alpha}_{-4}, \boldsymbol{\alpha}_{-2}, \boldsymbol{\alpha}_0$ is zero at infinity, and the stress field correspond to $\boldsymbol{\alpha}_1$ is a constant matrix:

$$\sigma_1 = \frac{1}{2}\sqrt{\frac{3}{2\pi}} \begin{pmatrix} 0 & 0 & 0 & \sqrt{2}(-1+2v^j) & 1-2v^j & 0 & 2(-2+v^j) & 0 & 0 \\ 0 & 0 & 0 & \sqrt{2}(-1+2v^j) & 2(-2+v^j) & 0 & 1-2v^j & 0 & 0 \\ 0 & 0 & 0 & -2\sqrt{2}(-2+v^j) & 1-2v^j & 0 & 1-2v^j & 0 & 0 \\ 0 & 0 & 0 & 0 & 0 & 0 & 0 & -\frac{5}{2}+2v^j & 0 \\ 0 & 0 & 0 & 0 & 0 & -\frac{5}{2}+2v^j & 0 & 0 & 0 \\ 0 & 0 & 0 & 0 & 0 & 0 & 0 & 0 & -\frac{5}{2}+2v^j \end{pmatrix} \quad \text{(B.5)}$$

From Eq.(B.5), we can clearly see that the 1$^{\text{st}}$ order P-N solutions correspond to the 6 independent constant-stress modes, in addition to 3 rigid-body modes. Thus, we can easily determine the coefficients for $\boldsymbol{\alpha}_1$ by satisfying the far-field boundary condition given in Eq.(4).

The other coefficients in Eq.(B.4) can be determined by enforcing the interface condition described by Eq.(12) and Eq.(14). This procedure can be easily completed employing the symbolic computation tool Mathematica, and it only takes a few minutes. After the coefficients are determined, we substitute them back into Eq.(B.4) to obtain the expression in Eqs.(27-32).

nano-scale spherical inclusion due to interface stress. International Journal of Solids and Structures 43, 5055-5065.

Lu, T.Q., Zhang, W.X., Wang, T.J., 2011. The surface effect on the strain energy release rate of buckling delamination in thin film-substrate systems. International Journal of Engineering Science 49, 967-975.

Lurie, A.I., 2005. Three-dimensional problems in the theory of elasticity, Theory of Elasticity. Springer Berlin Heidelberg, Berlin, Heidelberg, pp. 243-407.

Manav, M., Anilkumar, P., Phani, A.S., 2018. Mechanics of polymer brush based soft active materials– theory and experiments. Journal of the Mechanics and Physics of Solids 121, 296-312.

McDowell, M.T., Leach, A.M., Gall, K., 2008. Bending and tensile deformation of metallic nanowires. Modelling and Simulation in Materials Science and engineering 16, 045003.

Medasani, B., Park, Y.H., Vasiliev, I., 2007. Theoretical study of the surface energy, stress, and lattice contraction of silver nanoparticles. Physical Review B 75.

Mi, C., 2018. Elastic behavior of a half-space with a Steigmann–Ogden boundary under nanoscale frictionless patch loads. International Journal of Engineering Science 129, 129-144.

Mi, C., Kouris, D., 2014. On the significance of coherent interface effects for embedded nanoparticles. Mathematics and Mechanics of Solids 19, 350-368.

Miller, R.E., Shenoy, V.B., 2000a. Size-dependent elastic properties of nanosized structural elements. Nanotechnology 11, 139-147.

Miller, R.E., Shenoy, V.B., 2000b. Size-dependent elastic properties of nanosized structural elements. Nanotechnology 11, 139.

Nazarenko, L., Stolarski, H., Altenbach, H., 2016. Effective properties of short-fiber composites with Gurtin-Murdoch model of interphase. International Journal of Solids and Structures 97-98, 75-88.

Neuber, H.v., 1934. Ein neuer ansatz zur lösung räumlicher probleme der elastizitätstheorie. der hohlkegel unter einzellast als beispiel. ZAMM-Journal of Applied Mathematics and Mechanics/Zeitschrift für Angewandte Mathematik und Mechanik 14, 203-212.

Ogden, R.W., Steigmann, D.J., Haughton, D.M., 1997. The Effect of Elastic Surface Coating on the Finite Deformation and Bifurcation of a Pressurized Circular Annulus. J Elast 47, 121-145.

Papkovich, P., 1932. Solution générale des équations differentielles fondamentales d'élasticité exprimée par trois fonctions harmoniques. CR Acad. Sci. Paris 195, 513-515.

Rouhi, H., Ansari, R., Darvizeh, M., 2016. Size-dependent free vibration analysis of nanoshells based on the surface stress elasticity. Applied Mathematical Modelling 40, 3128-3140.

Sahmani, S., Aghdam, M.M., Akbarzadeh, A.H., 2016. Size-dependent buckling and postbuckling behavior of piezoelectric cylindrical nanoshells subjected to compression and electrical load. Materials and Design 105, 341-351.

Sharma, P., Ganti, S., Bhate, N., 2003. Effect of surfaces on the size-dependent elastic